# Generation of Bright Isolated Attosecond Soft X-Ray Pulses Driven by Multi-Cycle Mid-Infrared Lasers

M.-C. Chen[1,2], C. Hernández-García[1,3], C. Mancuso[1], F. Dollar[1], B. Galloway[1], D. Popmintchev[1], P.-C. Huang[2], B. Walker[4], L. Plaja[3], A. Jaron-Becker[1], A. Becker[1], T. Popmintchev[1], M. M. Murnane[1], H. C. Kapteyn[1]

[1]*JILA and Department of Physics, University of Colorado at Boulder, Boulder, Colorado 80309-0440, USA*

[2]*Institute of Photonics Technologies, National Tsing Hua University, Hsinchu 30013, Taiwan*

[3]*Grupo de Investigación en Óptica Extrema, Universidad de Salamanca, E-37008 Salamanca, Spain*

[4]*University of Delaware, Newark, DE 19716, USA*

*Communicating author: mingchang@mx.nthu.edu.tw*

**Abstract**

High harmonic generation driven by femtosecond lasers makes it possible to capture the fastest dynamics in molecules and materials. However, to date the shortest attosecond (*as*) pulses have been produced only in the extreme ultraviolet (EUV) region of the spectrum below 100 eV, which limits the range of materials and molecular systems that can be explored. Here we use advanced experiment and theory to demonstrate a remarkable convergence of physics: when mid-infrared lasers are used to drive the high harmonic generation process, the conditions for optimal bright soft X-ray generation naturally coincide with the generation of isolated attosecond pulses. The temporal window over which phase matching occurs shrinks rapidly with increasing driving laser wavelength, to the extent that bright isolated attosecond pulses are the norm for 2 μm driving lasers. Harnessing this realization, we demonstrate the generation of isolated soft X-ray attosecond pulses at photon energies up to 180 eV for the first time, that emerge as linearly chirped 300 *as* pulses with a transform limit of 35 *as*. Most surprisingly, we find that in contrast to *as* pulse generation in the EUV, long-duration, multi-cycle, driving laser pulses are required to generate isolated soft X-ray bursts efficiently, to mitigate group velocity walk-off between the laser and the X-ray fields that otherwise limit the conversion efficiency. Our work demonstrates a clear and straightforward approach for robustly generating bright attosecond pulses of electromagnetic radiation throughout the soft X-ray region of the spectrum.

**Introduction**

High-order harmonic generation (HHG) is the most extreme nonlinear optical process in nature, making it possible to coherently upconvert intense femtosecond laser light to much shorter wavelengths (*1*, *2*). HHG emission emerges as a broad spectrum with attosecond temporal structure that has made it possible to directly access the fastest time scales relevant to electron dynamics in atoms, molecules and materials. The unique properties of attosecond HHG in the extreme ultraviolet (EUV) have uncovered new understanding of fundamental processes in atoms, molecules, plasmas and materials, including the timescales on which electrons are emitted from atoms (*3*), the timescale for spin-spin and electron-electron interactions (*4*, *5*), the timescale that determines molecular dissociation and electron localization (*6-9*), the timescale and mechanisms for spin and energy transport in nanosystems (*10-12*), as well new capabilities to implement EUV microscopes with wavelength-limited spatial resolution (*13*).

Using multi-cycle 0.8 μm driving lasers, HHG generally emerges as a train of *as* pulses (*14*, *15*), narrowing to a single isolated *as* burst when the driving laser field is ~5 fs in duration (*16*, *17*). Other techniques can isolate a single burst using a combination of multicolor fields and polarization control (*18-24*) or spatial lighthouse gating of the driving laser pulses (*25*, *26*). Ionization and phase-matching gating within the generating medium can also result in bright isolated *as* pulse generation for short driving laser pulses (*27*, *28*). The chirp present on attosecond bursts can be compensated by using thin materials, gases, or chirped mirrors (*29-31*). To date however, most schemes for creating isolated attosecond pulses require the use of very short-duration few-cycle 0.8 μm driving laser pulses that are difficult to reliably generate or complex polarization modulation schemes. In addition, the carrier envelope phase (CEP) of the driving laser pulse must be stabilized.

A more general understanding of how to efficiently sculpt the temporal, spatial and spectral characteristics of HHG emission over an extremely broad photon energy range (from the EUV to the keV and higher) has emerged in recent years. This understanding is critical both for a fundamental understanding of strong-field quantum physics, as well as for applications which have fundamentally different needs in terms of the HHG pulse duration, spectral bandwidth and flux. By considering both the microscopic single-atom response as well as the macroscopic coherent build-up of HHG, efficient phase-matched HHG can now be implemented from the EUV to > keV photon energies, simply by driving HHG with mid-infrared (mid-IR) driving lasers (*32-36*). This advance represents the first general purpose, tabletop, coherent x-ray light source (*36*).

In this paper, we demonstrate a beautiful convergence of physics for mid-infrared (2 μm) driving lasers by showing that the conditions for optimal bright soft X-ray generation naturally coincide with the generation of bright, isolated, attosecond soft X-ray bursts. We combine advanced theory with a novel experimental method equivalent to high resolution Fourier Transform spectroscopy to experimentally measure bright, attosecond, soft X-ray pulses for the first time - in this case linearly chirped, 300 as pulses with a transform limit of 35 as, extending to photon energies around 180 eV. We also show that the temporal window during which phase matching occurs shrinks rapidly with increasing driving laser wavelength, Most surprisingly, we show that optimal bright attosecond pulse generation in the soft X-ray region *requires* the use of longer-duration, multi-cycle, mid-IR driving lasers, to mitigate group velocity walk-off issues that would otherwise reduce the conversion efficiency. By harnessing the beautiful physics of phase matching, this work represents the simplest and most robust scheme for attosecond soft X-ray pulse generation, and will make attosecond science and technology accessible to a broader community.

**Experiment**

In our experiment, laser pulses at wavelengths of 0.8 μm, 1.3 μm, and 2.0 μm are generated using a 1kHz Ti:sapphire laser pumping a three-stage optical parametric amplifier (OPA). The pulse durations at all three wavelengths were adjusted to be ≈ 10 cycles in duration (24 fs at 0.8 μm – 9.5 cycles, 35 fs at 1.3 μm – 8 cycles, and 90 fs at 2.0 μm – 13.5 cycles) and were measured using second-harmonic frequency resolved optical gating. Each driving laser beam was focused into a 2-mm-long, Ar-filled cell, as shown in Fig. 1. The laser peak intensity at the focus was estimated by measuring the maximum HHG photon energy and then using the single atom cutoff rule ($I \cdot \lambda_L^2$) (*40, 41*). The lens was translated to precisely control the focal position and obtain optimal phase matching (*42*).

The generated HHG beam was refocused using a Kirkpatrick-Baez (KB) mirror pair. Two replicas of the HHG beam were then obtained by using an in-line, partially-split-mirror, spatial beam separator that can delay a part of the HHG beam with respect to itself with ultrahigh temporal resolution of ≈ 1.5 as. This enabled a precise field autocorrelation of the HHG emission to be made by monitoring the resultant fringes using an X-ray CCD (Andor). The HHG spectra obtained from two different methods were then compared. First, a fast Fourier transform (FFT) of the field autocorrelation trace was performed (see Fig. 2). Second, the HHG spectra were measured independently using a flat-field imaging soft x-ray spectrometer, which was calibrated using the absorption edges of several thin metal filters.

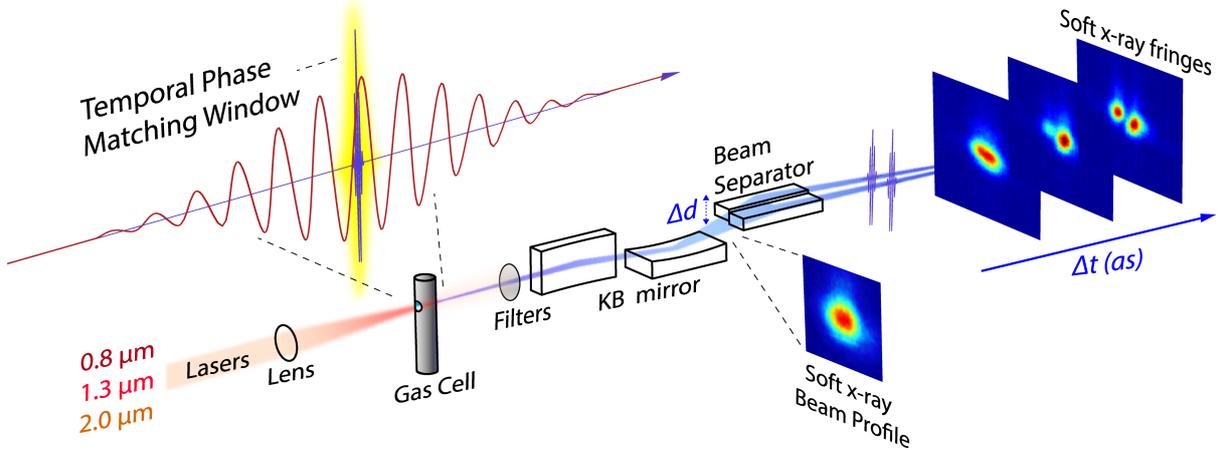

**Figure 1**: Schematic of the setup for attosecond high resolution Fourier Transform spectroscopy. The number of discrete bursts in an attosecond pulse train driven by 0.8 μm, 1.3 μm and 2.0 μm lasers is *directly* measured using soft x-ray interferometry, by delaying one part of the HHG pulse with respect to itself with ultrahigh precision of 1.5 *as*.

Figure 2 plots the experimental HHG field autocorrelations and their corresponding Fourier transforms at low and high laser intensity conditions for 0.8 μm, 1.3 μm and 2 μm 10-cycle FWHM laser fields. (High laser intensity corresponds to optimal phase-matching at the highest photon energies possible in Ar, generating the brightest HHG flux.) When driving the HHG process with a 0.8 μm laser, we observe that the number of X-ray bursts decreases (from 15 to 9) as the driving laser intensity is increased (from $1.5 \times 10^{14}$ to $2.6 \times 10^{14}$ W/cm$^2$). For the case of a 1.3 μm driving laser, the number of individual bursts decreases faster (from 9 to 4 bursts as the laser intensity is increased from $1.3 \times 10^{14}$ to $2.1 \times 10^{14}$ W/cm$^2$). Finally, for a 2 μm driving laser, we obtain a remarkable result: the number of individual harmonic bursts decreases to a single isolated burst as the driving laser intensity reaches $1.6 \times 10^{14}$ W/cm$^2$. Moreover, the HHG flux increases with increasing laser intensity. For 2 $\mu$m driving lasers in particular, the HHG emission shifts to a central photon energy of 140 eV (spanning from 90 to 180 eV), with a FWHM bandwidth of ≈ 60 eV capable of supporting a ≈ 35 *as* transform-limited pulse. This pulse duration is corroborated by the measured 70 *as* field autocorrelation trace, that as expected, corresponds to twice the pulse duration. As discussed in more detail below, although the single isolated burst is linearly chirped to ≈ 300 *as* duration, it nevertheless represents the highest photon energy, broadest bandwidth, isolated *as* pulse characterized to date (*24*). Moreover, both the number of individual HHG bursts and the spectrum are insensitive to the phase of the laser carrier wave with respect to the pulse envelope (CEP), which was verified by adjusting the CEP of the 2 μm laser, as suggested by earlier experiments using 0.8 $\mu$m driving lasers (*28*, *43*, *44*).

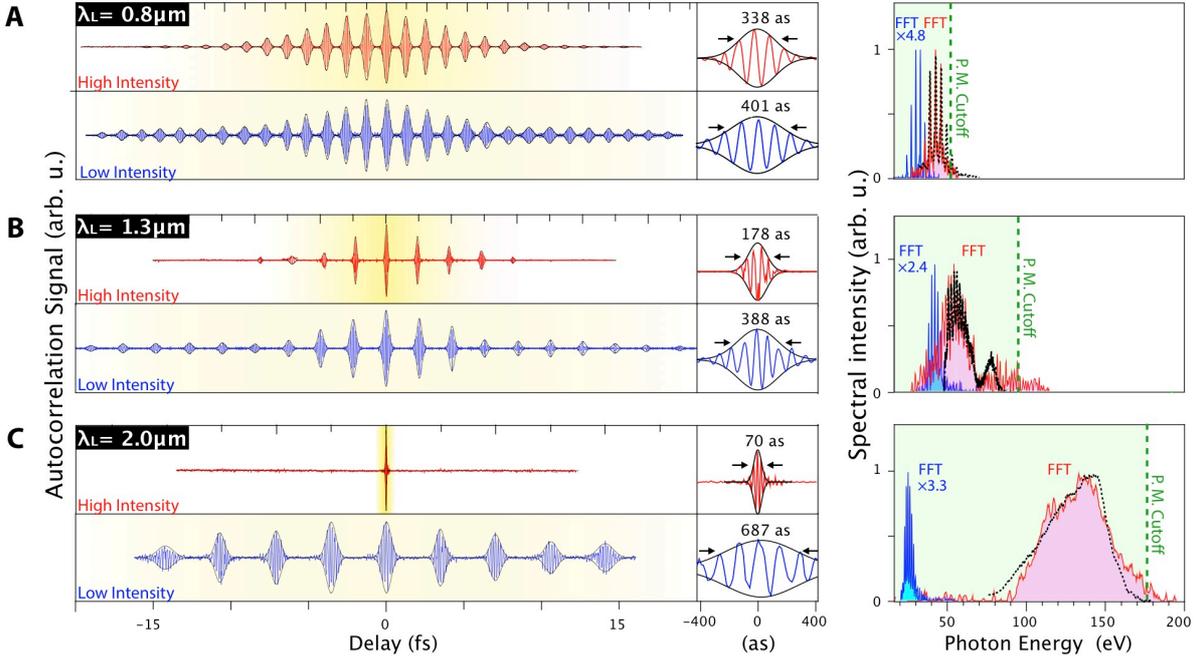

**Figure 2**: Comparison of the experimental HHG autocorrelation data (normalized) from Ar driven by 10 cycle laser pulses at central wavelengths of (A) 0.8μm, (B) 1.3 μm, and (C) 2 μm for high and low laser intensity conditions [Red and Blue lines]. Left: field autocorrelation of the HHG field and enlarged view near time zero with the coherence time of the central pulse envelope. Note that the bandwidth-limited pulse duration is *half* of this coherence time. The temporal window during which phase matching is possible is highlighted in yellow. Right: HHG spectra obtained from the fast Fourier transform of the field autocorrelation traces (filled-area plots), showing excellent agreement with the experimental spectra (black dotted lines). The predicted phase matching cutoffs are also shown [greed dashed lines].

The high laser intensity used to obtain the data plotted in Fig. 2 is selected to access the phase matching photon energy limits in Ar for each driving laser wavelength for 10-cycle laser pulses (≈ 50 eV for 0.8 $\mu$m, 95 eV for 1.3 $\mu$m, and 175 eV for 2 $\mu$m). As discussed in detail below, phase matching is supported only during a narrow temporal window within the laser pulse. On the leading edge of the laser pulse, the laser phase velocity is < c (speed of light), due to the dominance of neutral atoms. In contrast, on the trailing edge, the laser phase velocity is > c due to ionization of the gas beyond the critical ionization level, which terminates phase matching (*32*). In the case of mid-IR driving lasers, three factors lead to a much shorter phase matching window when compared with 0.8 $\mu$m lasers: the harmonic order is higher, the period of the driving wave is longer, and the phase matching pressure and ionization level are higher. Each of these factors contributes to a larger phase shift within each half-cycle and, therefore, to a shorter phase matching window (see Fig. 3). In contrast, for laser intensities below the critical ionization limit, optimal pressure and ionization are low, the harmonic order is low, and phase matching extends over many laser cycles. The practical consequence of this physical scaling is that for *any driving laser wavelength,* the temporal phase matching window can be narrowed down by increasing the driving laser intensity and gas pressure. However, for longer driving laser wavelengths, the temporal narrowing effect can be much stronger, so that isolated attosecond pulse generation is more robust and simple.

The flux of the HHG beam initially grows quadratically with pressure, but saturates at different pressures depending on the driving laser wavelengths and the absorption lengths ($L_{abs}$) in various regions of the spectrum. The general trend is that higher HHG photon energies require longer-wavelength mid-IR driving lasers and significantly higher pressure-length products for optimal conversion efficiency (*32*). Using a 0.8 μm driving laser, the brightest (phase matched) HHG emission peaks at 40 eV where the optimal backing pressure is 80 torr. However, for 1.3 and 2 $\mu$m driving lasers, the phase matched peaks shift to 65 eV at 350 torr, and 140 eV at 600 torr, respectively. These optimized phase matching pressures, observed experimentally, are all in excellent agreement with the predicted absorption-limited HHG emission for media lengths $L_{med} \sim 3L_{abs}$ ($L_{med}$ = 2 mm; for Ar, $L_{abs}$ = 0.81 mm at 40 eV, 0.67 mm at 65 eV, and 0.59 mm at 140 eV) (*45, 46*).

We emphasize that the attosecond field autocorrelation demonstrated here is sufficient to fully understand and characterize how temporal gating of phase matching scales with driving laser wavelength, provided that the emission indeed exhibits coherence in the time domain from burst-to-burst. Such an assumption is consistent with all experimental studies of HHG to-date (*15*). Moreover, the excellent experimentally-measured spatial coherence of phase matched HHG driven by mid-IR lasers also demonstrates temporal coherence across the entire beam (*34, 36, 47*). In contrast to attosecond streaking, the field autocorrelation trace can *directly* measure the number of attosecond bursts contained in the HHG emission, i.e., a total of *2n-1* fringes will be measured if there are *n* bursts in the pulse train, and this can be used for faster experimental feedback in optimization. The field autocorrelation approach is also significantly more rigorous than merely observing if the HHG spectrum measured using a grating spectrometer forms a discrete or continuous spectrum. A supercontinuum spectral structure does not necessarily imply the presence of an isolated X-ray burst, because the attainable resolving power of a grating spectrometer is strongly dependent on the number of grating groove illuminated, the smallest slit width used, and aberrations induced by the optical elements in the system. In contrast, the FFT-generated spectrum from a field autocorrelation trace can provide much higher resolving power, which is given by the Heisenberg's uncertainty relation of $\Delta E$ [eV]=4.1357/$T$ [fs], where $\Delta E$ is the FFT spectral resolution and $T$ represents the maximum temporal delay range. The delay range used in our HHG spatial beam separator was ≈ 300 fs, which supports an extremely high spectral resolution of ≈ 0.01 eV (much smaller than the energy spacing between individual harmonics of 0.62 eV). This capability for high-resolution spectral measurements is essential in order to resolve unambiguously whether the HHG spectrum is discrete or continuous in the case of a mid-IR driver, where the harmonic order can in some cases be as high as 5000 (*36*).

**Theory**

We use both an analytic theory and an advanced numerical model to understand why the phase matching window shrinks rapidly with increasing driving laser wavelength and how phase matching scales with the driving laser pulse duration in the mid-IR. In our analytical theory, we neglect variations in the geometric contribution to phase mismatch, which in experiment can be minimized either using a laser confocal parameter much longer than the medium length, or guide-wave configuration (*32*). Thus the major contributions to the phase mismatch $\Delta k = (k_{q\omega} - qk_\omega)$ are due to the pressure-dependent neutral and free-electron plasma dispersion terms that can be written as:

$$\Delta k(t) \approx P \cdot q \cdot \left([1 - \eta(t)] \cdot \delta n \cdot \frac{2\pi}{\lambda_L} - \eta(t) \cdot N_{atm} \cdot r_e \cdot \lambda_L\right), \quad (1)$$

where $P$ is the gas pressure, $q$ is the harmonic order, $\eta(t)$ is the instantaneous ionization fraction, $\delta n$ is the difference between the indices of refraction at the fundamental and harmonic wavelengths, $\lambda_L$ is the central wavelength of the driving laser, $N_{atm}$ is the number density of atoms at 1 atm, and $r_e$ is the classical electron radius. From Eq. 1, phase matching *($\Delta k = 0$)* will occur at some time t when $\eta(t) = \eta_c = [(N_{atm} r_e \lambda_L^2/\delta n\, 2\pi) + 1]^{-1}$ and $\eta_c$ is the critical ionization. If *$\Delta k \neq 0$*, the harmonic signal builds up over a shorter propagation distance, until the relative phase of the driving and harmonic fields are shifted by π radians. This defines the coherence length *$L_{coh}=\pi/\Delta k$*, or the distance over which the HHG fields add constructively.

Figure 3 shows plots of the phase mismatch and coherence length as a function of time during the laser pulse for different driving laser wavelengths (based on Eq. 1), and assuming Ammosov–Delone–Krainov (ADK) ionization rates (*48*). According to these predictions, the phase matching temporal window shrinks rapidly with increasing driving laser wavelength. This can be understood from Eq. 1: since *Δk* is proportional to the product of the pressure and harmonic order, in the mid-IR region of the spectrum the phase mismatch will be higher and the coherence length shorter for adjacent half-cycles near optimal phase matching. Efficient HHG emission also requires that the coherence length is larger than the medium length, and that the medium length is greater than the HHG reabsorption distance ($L_{abs}$) i.e. $L_{coh} > L_{med} > L_{abs}$ (*45, 46*). Under our experimental conditions, $L_{med} \sim 3L_{abs}$.

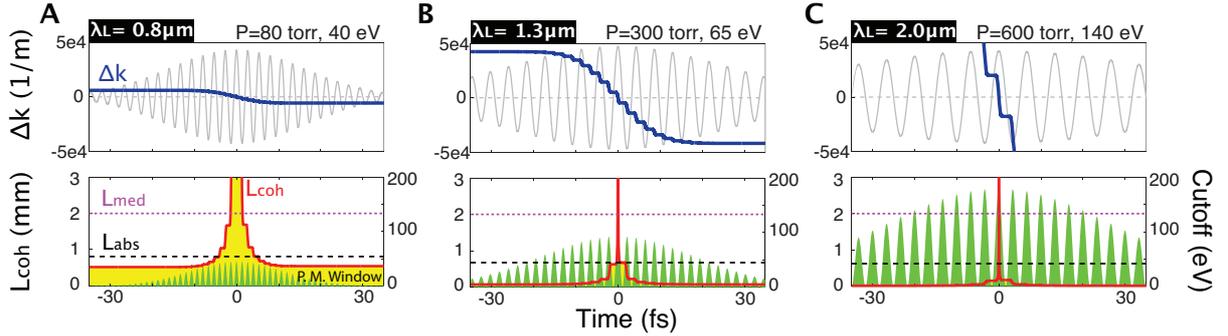

**Figure 3**: Calculated phase mismatch Δ*k* [blue line], $L_{coh}$ [red line and yellow highlight], and instantaneous HHG cutoff photon energy [green] for HHG in Ar driven by 10-cycle laser pulses: (A) 0.8 μm, 2.42 × 10$^{14}$ W/cm$^2$; (B) 1.3 μm, 1.87 × 10$^{14}$ W/cm$^2$; and (C) 2.0 μm, 1.5 × 10$^{14}$ W/cm$^2$.

Note that in Fig. 3, the laser intensity is adjusted so that perfect phase matching occurs at the peak of the laser pulse, to ensure the highest photon energies and brightest isolated attosecond pulse generation. Moreover, the pressure is increased for optimal phase matching so that emission from several absorption lengths of the HHG light is possible. In general, the position of the phase-matching window along the laser pulse depends on both the laser intensity and wavelength, whereas the width of the phase-matching window depends on the pressure-length product. This temporal width determines the number of individual bursts in the attosecond pulse train, as well as the HHG conversion efficiency. As the driving laser intensity is increased further, the phase matching window moves along the pulse front, while both the phase matching energy and the width of the phase matching window are nearly constant.

We can also analytically estimate the width of the phase matching window by calculating the time dependence of phase mismatch Δ*k* from Eq. 1:

$$\frac{\partial \Delta k(\tau)}{\partial \tau} \approx -P \cdot q \cdot \left[ \delta n \cdot \frac{2\pi}{\lambda_L} + N_{atm} \cdot r_e \cdot \lambda_L \right] \cdot \frac{d\eta(\tau)}{d\tau}$$

$$\propto P \cdot q \cdot \lambda_L \cdot \frac{d\eta(\tau)}{d\tau}, \qquad (2)$$

where $\tau$ represents time measured in cycles of the driving laser. Note the neutral gas dispersion contribution in Eq. 2 is negligible compared to the free electron dispersion contribution, especially for long wavelength driving lasers. At low laser intensities, the harmonic order $q$ and $\frac{d\eta(\tau)}{d\tau}$ are small, which suppresses phase matching gating, in agreement with the data of Fig. 2 (low intensity autocorrelation). Near optimal phase-matching, the pressure, $P_{PM}$, and the central harmonic order, $q$, scale by the laser wavelength as $P_{PM} \propto \lambda_L^2$, and $q \propto 1/2 \cdot \lambda_L^{2.7}$, as have been observed experimentally (*32, 34, 36*). The scaling of $q$ arises from two contributions; a factor of $1/2 \cdot \lambda_L^{1.7}$ since the central energy of the *as* burst is half the phase-matching cutoff (*32*), and an additional factor of $\lambda_L$ from the fundamental laser photon energy. Thus, under optimal phase-matching condition, Eq. 2 can be approximated by $\frac{\partial \Delta k(\tau)}{\partial \tau} \propto \lambda_L^{5.7} \cdot \frac{d\eta(\tau)}{d\tau}$, which scales strongly with the wavelength of driving laser. The combined effects of higher pressures needed for bright HHG emission, the resulting higher harmonic orders, and a stronger free electron dispersion for long wavelength driving lasers creates a large phase mismatch jump between adjacent half-cycles of the driving laser. Note that the larger separation of the half-cycles of the driving laser further isolates the *as* bursts. Consequently, a longer-wavelength driving laser can much more easily induce strong phase matching temporal gating (see Fig. 2) i.e. a smaller perturbation away from optimal phase matching conditions – for example by increasing laser intensity or gas pressure – will isolate a single attosecond soft X-ray burst more easily when using mid-IR driving lasers compared with Ti:sapphire driving lasers (*28, 43, 44*).

We also performed advanced 3D numerical macroscopic HHG simulations, using the method presented in (*38*). We discretize the gas medium into elementary radiators, and propagate the radiation of each of these sources to the detector,

$$E_j(r_d, t) = \frac{q_j s_d}{c^2 |r_d - r_j|} \left[ s_d \times a_j \left( t - \frac{|r_d - r_j|}{c} \right) \right], \qquad (3)$$

where $q_j$ is the charge of the electron, $s_d$ is the unitary vector pointing to the detector, and $r_d$ and $r_j$ are the position vectors of the detector and of the elementary radiator, respectively. The dipole acceleration $a_j$ of each elementary source is computed using the SFA+ method, which is an extension of the standard strong field approximation (*39*). The signal at the detector is computed as the coherent addition of the HHG contributions of all the elementary sources where the HHG light is assumed to propagate to the detector with a phase velocity $c$. Propagation effects of the fundamental field, including plasma and neutral dispersion as well as time-dependent group velocity walk-off, are all taken into account. The absorption of the harmonics in the gas is modeled using Beer's law, through a 2 mm Argon gas cell of uniform density, at different pressures. The laser pulse was modeled in time as a Gaussian envelope, 8 cycle FWHM (53 fs), with peak intensity of $1.2 \times 10^{14}$ W/cm$^2$, chosen to reach optimal phase-matching conditions ($\Delta k = 0$) at the center of the 2 $\mu$m wavelength laser pulse.

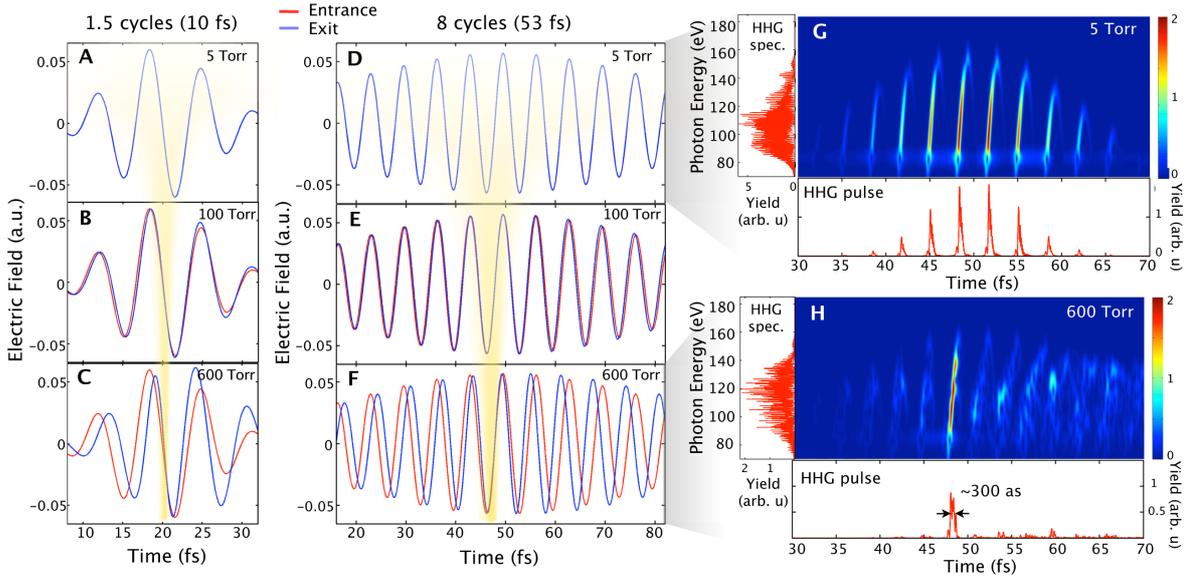

**Figure 4** Numerical calculations showing the dependence of the phase-matching window on the number of driving laser cycles and gas pressure. The laser field is shown both at the entrance (red) and exit (blue) of a 2 mm gas cell at pressures of (A, D) 5 torr, (B, E) 100 torr, and (C, F) 600 torr for 1.5 cycles and 8 cycles FWHM at phase-matching intensities of $1.3 \times 10^{14}$ W/cm$^2$ and $1.2 \times 10^{14}$ W/cm$^2$ respectively, from the central part of a driving laser field with a Bessel profile of radius 60 $\mu$m. The time window during which phase matching is possible is highlighted in yellow. The temporal profile of the HHG emission from the 8 cycles laser field is shown for (G) 5 torr and (H) 600 torr. The right-hand panels show a time-frequency analysis together with the HHG spectrum and temporal emission, which evolves from an *as* pulse train (G) to an isolated 300 *as* chirped pulse (H). Note that for a 1.5 cycle driving laser, at the high pressures required for bright HHG emission, phase matching is not possible due to group-velocity walk-off.

To illustrate strong phase matching temporal gating, as well as group velocity walk-off effects, Figs. 4A – C compare the HHG emission as a function of pressure for single cycle versus multi-cycle driving lasers pulses (1.5 cycle and 8 cycle pulses, respectively). The electric field of the laser is plotted at the entrance and the exit of the cell. As the pressure is increased from 5 to 600 torr, the increasing contribution of the neutral atom dispersion induces a phase-shift on the front of the pulse, while the presence of the free-electron plasma induces a chirp on the tailing edge of the pulse. The combined effect confines the phase-matching window to a sub-optical cycle duration (300 *as*) in the center of the pulse, where a bright, isolated, linearly chirped attosecond pulse (group delay dispersion ~0.005 fs$^2$) is generated at high pressures. As the 2 $\mu$m driving laser pulse duration is reduced from 8 to 1.5 cycles, there is a strong contribution due to group-velocity walk-off, that leads to a temporal delay of the envelope of the laser field as it propagates (see Fig. 4C). As a result, the phase matching window *completely disappears* for 1.5 cycle driving lasers at high pressures, thus inhibiting bright isolated *as* pulse generation. This group-velocity walk-off clearly is still present for an 8 cycle pulse, and the envelope of the pulse slightly lags the phase-matched half-cycle. However, near the peak of the pulse, this shift does not interfere with optimal phase-matching (Fig. 4D-F).

Finally, comparing HHG propagation for low and high pressures, for low pressures (5 torr), an attosecond pulse train is emitted, where the positive slope of the time-frequency structure (Fig. 4G) reveals the survival and phase matching of short trajectories after propagation. In the high-pressure case (600 torr), phase-matching temporal gating isolates a single attosecond burst from the previous train, over a time interval of ≈300 *as* (Fig. 4H). The neutral and free electron dispersion cause the attosecond HHG fields generated both at the front and back of the laser pulse to interfere destructively. The phase matched HHG bandwidth (≈ 60 eV FWHM) of the isolated attosecond pulse supports a Fourier transform limit of ~35 *as* temporal duration, which agrees very well with experimental observations.

In summary, we have found a robust method for generating bright isolated attosecond soft X-ray pulses by exploiting the time-dependent phase matching dynamics of high harmonic generation driven by multi-cycle mid-infrared lasers. Using a soft X-ray autocorrelator with 1.5 *as* precision to directly and precisely count the number of individual attosecond bursts, we provide the first unambiguous experimental evidence for generating isolated attosecond soft X-ray pulses. Surprisingly, we find that in contrast to *as* pulse generation in the EUV, long-duration, multi-cycle, driving laser pulses are required to generate isolated soft X-ray bursts efficiently, to mitigate group velocity walk-off between the laser and the X-ray fields that otherwise limit the conversion efficiency. Moreover, this approach doe not require the use of few-cycle driving lasers with carrier envelope stabilization or other complex technologies. This new technique presents an accessible and reliable route to generate stable, isolated attosecond pulses with pulse durations in the single digit attosecond and even zeptosecond range (*35, 36*). This method is applicable to all technologies for mid-IR ultrashort pulse generation, including emerging new laser materials, parametric amplification techniques, optical fiber or fiber-bulk lasers. Our results will expand the field of attosecond science, making it possible to capture the attosecond motion of electrons in a broader range of atoms, molecules, liquids and materials.

**Acknowledgements**

The authors acknowledge support from the DOE Office of Basic Energy Sciences (AMOS program) and used facilities provided by the NSF Center for EUV Science and Technology. M.-C. C. and P.-C. H. acknowledge support by Taiwan National Science Coucil (Grant No. 102WFA0400681). C.M. and B.G. gratefully acknowledge support from an NSF Graduate Research Fellowship and an NNSA SSGF Fellowship. C.H.-G. acknowledges support by a Marie Curie International Outgoing Fellowship within the EU Seventh Framework Programme for Research and Technological Development, under REA grant agreement No. 328334. L. P. acknowledges support from Junta de Castilla y León (Project No. SA116U13). A.J.-B.was supported by grants from the U.S. National Science Foundation (Awards No. PHY-1125844 and No. PHY-1068706). This work utilized the Janus supercomputer, which is supported by the National Science Foundation (Award No. CNS-0821794) and the University of Colorado Boulder. The Janus supercomputer is a joint effort of the University of Colorado Boulder, the University of Colorado Denver and the National Center for Atmospheric Research. Janus is operated by the University of Colorado, Boulder.

**References**

1. A. McPherson *et al.*, Studies of multiphoton production of vacuum-ultraviolet radiation in the rare gases, *J Opt Soc Am B* **4**, 595–601 (1987).
2. X. F. LI, A. L'Huillier, M. Ferray, L. A. LomprE, G. Mainfray, Multiple-Harmonic Generation in Rare-Gases at High Laser Intensity, *Phys. Rev. A* **39**, 5751–5761 (1989).
3. K. Klünder *et al.*, Probing Single-Photon Ionization on the Attosecond Time Scale, *Phys. Rev. Lett.* **106**, 143002 (2011).
4. D. Rudolf *et al.*, Ultrafast magnetization enhancement in metallic multilayers driven by superdiffusive spin current, *Nature Communications* **3**, 1037–6 (1AD).
5. L. Miaja-Avila *et al.*, Direct Measurement of Core-Level Relaxation Dynamics on a Surface-Adsorbate System, *Phys. Rev. Lett.* **101**, 046101 (2008).
6. K. T. Kim *et al.*, Amplitude and Phase Reconstruction of Electron Wave Packets for Probing Ultrafast Photoionization Dynamics, *Phys. Rev. Lett.* **108**, 093001 (2012).
7. E. Gagnon *et al.*, Soft X-ray-Driven Femtosecond Molecular Dynamics, *Science* **317**, 1374–1378 (2007).
8. W. Li *et al.*, Visualizing electron rearrangement in space and time during the transition from a molecule to atoms, *Proceedings of the National Academy of Sciences* **107**, 20219–20222 (2010).
9. T. Rohwer *et al.*, Collapse of long-range charge order tracked by time-resolved photoemission at high momenta, *Nature* **471**, 490–493 (2012).
10. S. Mathias *et al.*, Probing the timescale of the exchange interaction in a ferromagnetic alloy, *Proceedings of the National Academy of Sciences* **109**, 4792–4797 (2012).

11. E. Turgut *et al.*, Controlling the Competition between Optically Induced Ultrafast Spin-Flip Scattering and Spin Transport in Magnetic Multilayers, *Phys. Rev. Lett.* **110**, 197201 (2013).
12. M. E. Siemens *et al.*, Quasi-ballistic thermal transport from nanoscale interfaces observed using ultrafast coherent soft X-ray beams, *Nature Materials* **9**, 26–30 (2009).
13. M. D. Seaberg *et al.*, Ultrahigh 22 nm resolution coherent diffractive imaging using a desktop 13 nm high harmonic source, *Opt Express* **19**, 22470 (2011).
14. P. M. Paul *et al.*, Observation of a train of attosecond pulses from high harmonic generation, *Science* **292**, 1689–1692 (2001).
15. Y. Nabekawa, T. Shimizu, Y. Furukawa, E. Takahashi, K. Midorikawa, Interferometry of Attosecond Pulse Trains in the Extreme Ultraviolet Wavelength Region, *Phys. Rev. Lett.* **102**, 213904 (2009).
16. I. Christov, M. Murnane, H. Kapteyn, High-harmonic generation of attosecond pulses in the "single-cycle"" regime, *Phys. Rev. Lett.* **78**, 1251–1254 (1997).
17. E. Goulielmakis *et al.*, Single-Cycle Nonlinear Optics, *Science* **320**, 1614–1617 (2008).
18. P. B. Corkum, N. H. Burnett, M. Y. Ivanov, Subfemtosecond Pulses, *Opt Lett* **19**, 1870–1872 (1994).
19. I. J. Sola *et al.*, Controlling attosecond electron dynamics by phase-stabilized polarization gating, *Nature Physics* **2**, 319–322 (2006).
20. X. Feng *et al.*, Generation of isolated attosecond pulses with 20 to 28 femtosecond lasers, *Phys. Rev. Lett.* **103**, 183901 (2009).
21. C. Altucci *et al.*, Interplay between group-delay-dispersion-induced polarization gating and ionization to generate isolated attosecond pulses from multicycle lasers, *Opt Lett* **35**, 2798–2800.
22. E. J. Takahashi, P. Lan, O. D. Mücke, Y. Nabekawa, K. Midorikawa, Infrared two-color multicycle laser field synthesis for generating an intense attosecond pulse, *Phys. Rev. Lett.* **104**, 233901 (2010).
23. H. Mashiko *et al.*, Double Optical Gating of High-Order Harmonic Generation with Carrier-Envelope Phase Stabilized Lasers, *Phys. Rev. Lett.* **100**, 103906 (2008).
24. G. Sansone, L. Poletto, M. Nisoli, High-energy attosecond light sources, *Nat Photonics* **5**, 655–663 (2011).
25. J. A. Wheeler *et al.*, Attosecond lighthouses from plasma mirrors, *Nat Photonics* **6**, 829–833 (2012).
26. K. T. Kim *et al.*, Photonic streaking of attosecond pulse trains, *Nat Photonics* **7**, 651–656 (2013).
27. M. J. Abel *et al.*, Isolated attosecond pulses from ionization gating of high-harmonic emission, *Chem Phys* **366**, 9–14 (2009).
28. I. Thomann *et al.*, Characterizing isolated attosecond pulses from hollow-core waveguides using multi-cycle driving pulses, *Opt Express* **17**, 4611–4633 (2009).
29. R. López-Martens *et al.*, Amplitude and Phase Control of Attosecond Light Pulses, *Phys. Rev. Lett.* **94** (2005), doi:10.1103/PhysRevLett.94.033001.
30. K. Kim *et al.*, Self-Compression of Attosecond High-Order Harmonic Pulses, *Phys. Rev. Lett.* **99** (2007), doi:10.1103/PhysRevLett.99.223904.
31. A.-S. Morlens *et al.*, Compression of attosecond harmonic pulses by extreme-ultraviolet chirped mirrors, *Opt Lett* **30**, 1554–1556 (2005).
32. T. Popmintchev *et al.*, Phase matching of high harmonic generation in the soft and hard X-ray regions of the spectrum, *P Natl Acad Sci Usa* **106**, 10516–10521 (2009).
33. Popmintchev, T., Chen, M.-C., Bahabad, A., Murnane, M. M. & Kapteyn, H. C. Phase-matched Generation of Coherent Soft and Hard X-rays Using IR Lasers. *Provisional United States Patent Application:* 61171783 (2008), *United States Patent:* 8,462,824 (2013).
34. M.-C. Chen *et al.*, Bright, coherent, ultrafast soft x-ray harmonics spanning the water window from a tabletop light source, *Phys. Rev. Lett.* **105**, 173901 (2010).
35. T. Popmintchev, M.-C. Chen, P. Arpin, M. M. Murnane, H. C. Kapteyn, The attosecond nonlinear optics of bright coherent X-ray generation, *Nat Photonics* **4**, 822–832 (2010).
36. T. Popmintchev *et al.*, Bright Coherent Ultrahigh Harmonics in the keV X-ray Regime from Mid-Infrared Femtosecond Lasers, *Science* **336**, 1287–1291 (2012).
37. E. J. Takahashi, T. Kanai, K. L. Ishikawa, Y. Nabekawa, K. Midorikawa, Coherent water window X ray by phase-matched high-order harmonic generation in neutral media, *Phys. Rev. Lett.* **101**, 253901 (2008).
38. C. Hernandez-Garcia *et al.*, High-order harmonic propagation in gases within the discrete dipole approximation, *Phys. Rev. A* **82**, 033432 (2010).
39. J. A. Pérez-Hernández, L. Roso, L. Plaja, Harmonic generation beyond the Strong-Field Approximation:

the physics behind the short-wave-infrared scaling laws, *Opt Express* **17**, 9891–9903 (2009).

40. A. L'Huillier, K. Schafer, K. KULANDER, Theoretical Aspects of Intense Field Harmonic-Generation, *J. Phys. B: At. Mol. Opt. Phys.* **24**, 3315–3341 (1991).
41. P. B. Corkum, Plasma perspective on strong field multiphoton ionization, *Phys. Rev. Lett.* **71**, 1994 (1993).
42. P. Balcou, P. Salières, A. L'Huillier, Generalized phase-matching conditions for high harmonics: The role of field-gradient forces, *Phys. Rev. A* **55**, 3204–3210 (1997).
43. A. Sandhu *et al.*, Generation of sub-optical-cycle, carrier-envelope-phase—insensitive, extreme-uv pulses via nonlinear stabilization in a waveguide, *Phys. Rev. A* **74**, 061803 (2006).
44. S. Gilbertson, S. D. Khan, Y. Wu, M. Chini, Z. Chang, Isolated Attosecond Pulse Generation without the Need to Stabilize the Carrier-Envelope Phase of Driving Lasers, *Phys. Rev. Lett.* **105**, 093902 (2010).
45. E. Constant *et al.*, Optimizing high harmonic generation in absorbing gases: Model and experiment, *Phys. Rev. Lett.* **82**, 1668–1671 (1999).
46. C. G. Durfee III *et al.*, Phase matching of high-order harmonics in hollow waveguides, *Phys. Rev. Lett.* **83**, 2187 (1999).
47. R. Bartels *et al.*, Generation of spatially coherent light at extreme ultraviolet wavelengths, *Science* **297**, 376–378 (2002).
48. M. Ammosov, N. Delone, V. Krainov, Tunnel ionization of complex atoms and of atomic ions in an alternating electromagnetic field, *Sov Phys JETP* **64**, 1191–1194 (1986).